\documentclass[12pt]{article}
\usepackage{amssymb,amsmath,epsfig}

\begin{document}

\title{\bf $F(T)$ Models within Bianchi Type $I$ Universe}
\author{M. Sharif \thanks {msharif@math.pu.edu.pk} and Shamaila
Rani \thanks{shamailatoor.math@yahoo.com}\\
Department of Mathematics, University of the Punjab,\\
Quaid-e-Azam Campus, Lahore-54590, Pakistan.}

\date{}
\maketitle

\begin{abstract}
In this paper, we consider spatially homogenous and anisotropic
Bianchi type $I$ universe in the context of $F(T)$ gravity. We
construct some corresponding models using conservation equation and
equation of state parameter representing different phases of the
universe. In particular, we take matter dominated era, radiation
dominated era, present dark energy phase and their combinations. It
is found that one of the models has a constant solution which may
correspond to the cosmological constant. We also derive equation of
state parameter by using two well-known $F(T)$ models and discuss
cosmic acceleration.
\end{abstract}

\section{Introduction}

Modified theories of gravity have recently gained a lot of
interest due to its possible explanation about dark energy (DE).
Modern cosmology is in a state of crises in a sense that it
started with dark matter and erupted with an indication that most
of the universe is made up of DE. The discovery of the accelerated
expansion of the universe \cite{1}-\cite{S2} indicates that the
universe is nearly spatially flat and consists of about $74\%$ DE
causing cosmic acceleration. This unknown energy (having negative
pressure) is physically equivalent to vacuum energy and is almost
equally distributed in the universe. It has been used as an
essential factor in a recent attempt to formulate a cyclic model
of the universe. The universe (in its developing process) passes
transiently through the stiff fluid era $(\omega=1)$, the
radiation dominated era $(\omega=1/3)$, matter dominated era
$(\omega=0)$, transition era $(\omega=-1/3)$ and then tends to the
DE dominated era $(\omega=-1)$. In General Relativity (GR), the
simplest and most appealing candidate of DE is the cosmological
constant. However, it suffers from two serious theoretical
problems, the cosmological constant problem and coincidence
problem \cite{2}.

In alternative theories, $f(R)$ theory has many applications in
cosmology and gravity. This theory can directly be achieved by
replacing the Ricci scalar $R$ by $f(R)$ in the Einstein-Hilbert
action in GR. The study of the physics of $f(R)$ models is,
however, hampered by the complexity of the fourth order field
equations in the framework of metric formalism
\cite{S3}-\cite{S12}. Following the same scheme of modification in
action, $F(T)$ gravity can be obtained by replacing torsion scalar
$T$ by its general function $F(T)$ in the Lagrangian of
teleparallel gravity \cite{S13}-\cite{S19}. It helps to explain
the accelerated universe without introducing any DE component
\cite{4}-\cite{7}.

The $F(T)$ gravity models use the Weitzenb$\ddot{o}$k connection
which has no curvature but only torsion. Here torsion is responsible
for the accelerated expansion of universe and is formed using four
parallel vector fields, called vierbiens \cite{8}, which are
linearly independent. In this framework, the torsion tensor is
formed from the products of the first derivatives of tetrad. An
important advantage of this theory is that its field equations are
of second order and hence easy to tackle as compared to $f(R)$
theory.

Myrzakulov \cite{9} discussed different $F(T)$ models including
scalar fields and gave analytical solutions for the scale factors
and scalar fields. The same author studied the relationship between
$F(T)$ gravity and $k$-essence \cite{10} and also presented some new
models of purely kinetic $k$-essence. Karami and Abdolmaleki
\cite{11} obtained equation of state (EoS) parameter of polytropic,
standard, generalized and modified Chaplygin gas in this modified
gravity scenario. Wu and Yu \cite{12} discussed the two new $F(T)$
models and showed how the crossing of phantom divide line takes
place. They also discussed the observational constraints
corresponding to these models. Dent et al. \cite{13} discussed
$F(T)$ cosmology both at background and perturbed level. They
derived expressions for growth factor, stability of this theory and
vector-tensor perturbations. Li et al. \cite{14} explored this
modified gravity and local Lorentz invariance and remarked that this
theory is not local Lorentz invariant. Yang \cite{15} introduced
some new $F(T)$ models and described their physical implications and
cosmological behavior. Bengochea \cite{16} investigated the
consequences of data sets in this modified gravity. All the above
mentioned work have been carried out for the FRW metric.

In this paper, we would reconstruct the $F(T)$ gravity models using
Bianchi type $I$ spacetime which is the generalization of FRW metric
\cite{17}. This theory becomes equivalent to GR if $F(T)$ is
replaced by a constant \cite{13}-\cite{14}. The paper is organized
as follows: In next section, we present some basics of teleparallel
gravity and the corresponding field equations for Bianchi $I$ are
given in section \textbf{3}. A detailed construction of $F(T)$
gravity models is given using two approaches in section \textbf{4}.
Section \textbf{5} is devoted to study the EoS parameter for two
particular models and also a discussion on cosmic acceleration is
provided. In the last section, we summarize and conclude the
results.

\section{Preliminaries}

In this section, we introduce briefly the teleparallel theory of
gravity and its generalization to $F(T)$ theory. The Lagrangian
density for teleparallel and $F(T)$ gravity are, respectively,
given as follows \cite{14}
\begin{eqnarray}
L_{T}&=&\frac{e}{16\pi G}T,\\\label{1} L_{F(T)}&=&\frac{e}{16\pi
G}F(T),
\end{eqnarray}
where $T$ is the torsion scalar, $F(T)$ is a general differentiable
function of torsion, $G$ is the gravitational constant and
$e=\sqrt{-g}$. Mathematically, the torsion scalar is defined as
\begin{equation}\label{3}
T=S_{\rho}~^{\mu\nu}T^{\rho}~_{\mu\nu},
\end{equation}
where $S_{\rho}~^{\mu\nu}$ and torsion tensor $T^{\rho}~_{\mu\nu}$
are given as follows
\begin{eqnarray}\label{4}
S_{\rho}~^{\mu\nu}&=&\frac{1}{2}(K^{\mu\nu}~_{\rho}
+\delta^{\mu}_{\rho}T^{\theta\nu}~_{\theta}-\delta^{\nu}_{\rho}T^{\theta\mu}~_{\theta}),\\
\label{6}T^{\lambda}~_{\mu\nu}&=&\Gamma^{\lambda}~_{\nu\mu}-
\Gamma^{\lambda}~_{\mu\nu}=h^{\lambda}_{i}
(\partial_{\mu}h^{i}_{\nu}-\partial_{\nu}h^{i}_{\mu}).
\end{eqnarray}
Here $h^{i}_{\mu}$ are the components of the non-trivial tetrad
field $h_{i}$ in the coordinate basis. It is an arbitrary choice to
choose the tetrad field related to the metric tensor $g_{\mu\nu}$ by
the following relation
\begin{equation}\label{4*}
g_{\mu\nu}=\eta_{ij}h_{\mu}^{i}h_{\nu}^{j},
\end{equation}
where $\eta_{ij}=diag(1,-1,-1,-1)$ is the Minkowski metric for the
tangent space. For a given metric there exist infinite different
tetrad fields $h^i_{\mu}$ which satisfy the following properties:
\begin{equation}\label{11*}
h^{i}_{\mu}h^{\mu}_{j}=\delta^{i}_{j},\quad
h^{i}_{\mu}h^{\nu}_{i}=\delta^{\nu}_{\mu}.
\end{equation}
The procedure for evaluating the tetrad field has been given in
many papers \cite{S13}-\cite{S19}. Notice that the Latin alphabets
$(i,j,...=0,1,2,3)$ will be used to denote the tangent space
indices and the Greek alphabets $(\mu,\nu,...=0,1,2,3)$ to denote
the spacetime indices. The contorsion tensor $K^{\mu\nu}~_\rho$ is
defined as
\begin{equation}\label{5}
K^{\mu\nu}~_{\rho}=-\frac{1}{2}(T^{\mu\nu}~_{\rho}
-T^{\nu\mu}~_{\rho}-T_{\rho}~^{\mu\nu})
\end{equation}
which is equal to the difference between Weitzenb$\ddot{o}$ck and
Levi-Civita connections. The variation of Eq.(\ref{1}) with
respect to the vierbein field leads to the following field
equations
\begin{equation}\label{2}
[e^{-1}\partial_{\mu}(eS_{i}~^{\mu\nu})
-h^{\lambda}_{i}T^{\rho}~_{\mu\lambda}S_{\rho}~^{\nu\mu}]F_{T}
+S_{i}~^{\mu\nu}\partial_{\mu}(T)
F_{TT}+\frac{1}{4}h^{\nu}_{i}F=\frac{1}{2}\kappa^{2}h^{\rho}_{i}T^{\nu}_{\rho}.
\end{equation}
Here $F_{T}=\frac{dF}{dT},~F_{TT}=\frac{d^{2}F}{dT^{2}},~
\kappa^{2}=8\pi G,~S_{i}~^{\mu\nu}=h^{\rho}_{i}S_{\rho}~^{\mu\nu}$
with antisymmetric property and $T_{\mu\nu}$ is the
energy-momentum tensor given as
\begin{equation}\label{*}
T^{\nu}_{\rho}=diag(\rho_m,-p_m,-p_m,-p_m),
\end{equation}
where $\rho_{m}$ is the density while $p_{m}$ is the pressure of
matter inside the universe.

\section{The Field Equations}

The line element for a flat, homogenous and anisotropic Bianchi type
$I$ universe is
\begin{equation}\label{8}
ds^{2}=dt^{2}-A^{2}(t)dx^{2}-B^{2}(t)dy^{2}-C^{2}(t)dz^{2},
\end{equation}
where the scale factors $A,~B$ and $C$ are functions of cosmic time
$t$ only. Using Eqs.(\ref{4*}) and (\ref{8}), we obtain the tetrad
components as follows \cite{18}
\begin{eqnarray}\label{9}
h^{i}_{\mu}=diag(1,A,B,C),\quad
h_{i}^{\mu}=diag(1,A^{-1},B^{-1},C^{-1})
\end{eqnarray}
which obviously satisfies Eq.(\ref{11*}). Substituting Eqs.(\ref{4})
and (\ref{6}) in (\ref{3}) and using (\ref{8}), it follows after
some manipulation
\begin{equation}\label{10}
T=-2\left(\frac{\dot{A}\dot{B}}{AB}+\frac{\dot{B}\dot{C}}{BC}+
\frac{\dot{C}\dot{A}}{CA}\right).
\end{equation}
The field equations (\ref{2}) for $i=0=\nu$ and $i=1=\nu$ turn out
to be
\begin{eqnarray}\label{11}
&&F-4\left(\frac{\dot{A}\dot{B}}{AB}+\frac{\dot{B}\dot{C}}{BC}
+\frac{\dot{C}\dot{A}}{CA}\right)F_{T}=2\kappa^{2}\rho_{m},\\\nonumber
&&2\left(\frac{\dot{A}\dot{B}}{AB}+2\frac{\dot{B}\dot{C}}{BC}
+\frac{\dot{C}\dot{A}}{CA}+\frac{\ddot{B}}{B}+\frac{\ddot{C}}{C}\right)F_{T}
-4\left(\frac{\dot{B}}{B}+\frac{\dot{C}}{C}\right)\\\nonumber
&&\times\left[\left(\frac{\ddot{A}}{A}
-\frac{\dot{A^{2}}}{A^{2}}\right)\left(\frac{\dot{B}}{B}+\frac{\dot{C}}{C}\right)
+\left(\frac{\ddot{B}}{B}-\frac{\dot{B^{2}}}{B^{2}}\right)\left(\frac{\dot{C}}{C}
+\frac{\dot{A}}{A}\right)\right.\\\label{12}
&&\left.+\left(\frac{\ddot{C}}{C}-\frac{\dot{C^{2}}}{C^{2}}\right)
\left(\frac{\dot{A}}{A}+\frac{\dot{B}}{B}\right)\right]F_{TT}-F=2\kappa^{2}p_{m}.
\end{eqnarray}
The conservation equation takes the form
\begin{equation}\label{13}
\dot{\rho_{m}}+\left(\frac{\dot{A}}{A}+\frac{\dot{B}}{B}+\frac{\dot{C}}{C}\right)
(\rho_{m}+p_{m})=0.
\end{equation}

The average scale factor $R$, the mean Hubble parameter $H$ and the
anisotropy parameter $\Delta$ of the expansion respectively become
\begin{eqnarray}\label{14}
R&=&(ABC)^{1/3},\\\label{15}
H&=&\frac{1}{3}\left(\frac{\dot{A}}{A}+\frac{\dot{B}}{B}+\frac{\dot{C}}{C}\right)
=\frac{\dot{R}}{R},\\\label{16*}
\Delta&=&\frac{1}{3}\sum_{i=1}^{3}(\frac{H_i}{H}-1)^2,
\end{eqnarray}
where $H_i$ are the directional parameters in the direction $x$,
$y$ and $z$ respectively given as
\begin{equation}\label{17*}
H_1=\frac{\dot{A}}{A},\quad H_2=\frac{\dot{B}}{B},\quad
H_3=\frac{\dot{C}}{C}.
\end{equation}
It is mentioned here that the isotropic expansion of the universe
is obtained for $\Delta = 0$ which further depends upon the values
of unknown scale factors and parameters involved in the
corresponding models \cite{S20}-\cite{S22}. Equation (\ref{10})
can be written as
\begin{equation}\label{T}
T=-9H^{2}+J,\quad
J=\frac{\dot{A^{2}}}{A^2}+\frac{\dot{B^{2}}}{B^2}+\frac{\dot{C^{2}}}{C^2}.
\end{equation}
implying that
\begin{equation}\label{T'}
H=\frac{1}{3}\sqrt{J-T}.
\end{equation}
If we take $F(T)=T$, then Eqs.(\ref{11}) and (\ref{12}) will
reduce to
\begin{eqnarray}\label{16}
&&\rho_{m}+\rho_{T}=\frac{1}{2\kappa^{2}}\left[{-4\left(\frac{\dot{A}\dot{B}}{AB}
+\frac{\dot{B}\dot{C}}{BC}+\frac{\dot{C}\dot{A}}{CA}\right)+T}\right],\\\label{17}
&&p_{m}+p_{T}=\frac{1}{2\kappa^2}\left[{2\left(\frac{\dot{A}\dot{B}}{AB}
+2\frac{\dot{B}\dot{C}}{BC}+\frac{\dot{C}\dot{A}}{CA}
+\frac{\ddot{B}}{B}+\frac{\ddot{C}}{C}\right)-T}\right],
\end{eqnarray}
where $\rho_{T},~p_{T}$ are the torsion contributions given by
\begin{eqnarray}\label{18}
\rho_{T}&=&\frac{1}{2\kappa^2}\left[-4\left(\frac{\dot{A}\dot{B}}{AB}+\frac{\dot{B}
\dot{C}}{BC}+\frac{\dot{C}\dot{A}}{CA}\right)\left(1-F_{T}\right)+T-F\right],\\\nonumber
p_{T}&=&\frac{1}{2\kappa^2}\left[2\left(\frac{\dot{A}\dot{B}}{AB}
+2\frac{\dot{B}\dot{C}}{BC}+\frac{\dot{C}\dot{A}}{CA}
+\frac{\ddot{B}}{B}+\frac{\ddot{C}}{C}\right)
(1-F_{T})\right.\\\nonumber&+&4\left(\frac{\dot{B}}{B}
+\frac{\dot{C}}{C}\right)\left\{\left(\frac{\ddot{A}}{A}-\frac{\dot{A^{2}}}{A^{2}}\right)
\left(\frac{\dot{B}}{B}+\frac{\dot{C}}{C}\right)+\left(\frac{\ddot{B}}{B}-
\frac{\dot{B^{2}}}{B^{2}}\right)\right.\\\label{19}&\times&\left.\left.
\left(\frac{\dot{C}}{C} +\frac{\dot{A}}{A}\right)
+\left(\frac{\ddot{C}}{C}-\frac{\dot{C}^2}
{C^{2}}\right)\left(\frac{\dot{A}}{A}+\frac{\dot{B}}{B}\right)\right\}F_{TT}-T+F\right].
\end{eqnarray}

The relationship between energy density $\rho$ and pressure of
matter $p$ is described by EoS, $p=\omega\rho$, where $\omega$ is
the EoS parameter. For normal, relativistic and non-relativistic
matters, EoS parameter has different corresponding values. Using
Eqs.(\ref{11}) and (\ref{12}), the EoS parameter is obtained as
follows
\begin{equation}\label{20}
\omega=-1+\frac{(4Y-2E)F_{T}-4ZF_{TT}}{-4UF_{T}+F},
\end{equation}
where
\begin{eqnarray}\label{21}
E&=&\frac{\dot{A}\dot{B}}{AB}+2\frac{\dot{B}\dot{C}}{BC}+\frac{\dot{C}
\dot{A}}{CA}+\frac{\ddot{B}}{B}+\frac{\ddot{C}}{C}, \\\label{22}
Y&=&\frac{\dot{B}\dot{C}}{BC}+\frac{\ddot{B}}{B}+\frac{\ddot{C}}{C},\\\nonumber
Z&=&\left(\frac{\dot{B}}{B}
+\frac{\dot{C}}{C}\right)\left[\left(\frac{\ddot{A}}{A}-\frac{\dot{A^{2}}}{A^{2}}\right)
\left(\frac{\dot{B}}{B}+\frac{\dot{C}}{C}\right)+\left(\frac{\ddot{B}}{B}-
\frac{\dot{B^{2}}}{B^{2}}\right)\right.\\\label{23}&\times&\left.
\left(\frac{\dot{C}}{C} +\frac{\dot{A}}{A}\right)
+\left(\frac{\ddot{C}}{C}-\frac{\dot{C}^2}
{C^{2}}\right)\left(\frac{\dot{A}}{A}+\frac{\dot{B}}{B}\right)\right],\\\label{24}
U&=&\frac{\dot{A}\dot{B}}{AB}+\frac{\dot{B}\dot{C}}{BC}+\frac{\dot{C}\dot{A}}{CA}.
\end{eqnarray}
It is mentioned here that the homogenous part of Eq.(\ref{11})
yields the following solution
\begin{equation}\label{27}
F(T)=\frac{c_{0}}{\sqrt{T}},
\end{equation}
where $c_{0}$ is an integration constant. Using this equation in
Eq.(\ref{12}), we obtain
\begin{equation}\label{28}
p_{m}=\frac{1}{2\kappa^{2}}\left(\frac{3M\dot{T}}{2T^2}-\frac{3\dot{H}+J+L}{T}-\frac{1}{2}\right)
\frac{c_{0}}{\sqrt{T}}.
\end{equation}
where $L=\frac{\dot{B}\dot{C}}{BC}-\frac{\ddot{A}}{A}$ and
$M=\frac{\dot{B}}{B}+\frac{\dot{C}}{C}$. We would like to mention
here that $p_m$ vanishes for the FRW spacetime \cite{4}.

\section{Construction of Some $F(T)$ Models}

Here we construct some $F(T)$ models with different cases of perfect
fluid by using two approaches. In the first approach, we use the
continuity equation (\ref{13}) while in the second approach, EoS
parameter (\ref{20}) will be used. As the constituents of the
universe are non-relativistic matter, radiations and DE, we consider
the corresponding values of $\omega$ in the following subsections.

\subsection{Using Continuity Equation}

In this approach, we use the following relation \cite{19} for the
Bianchi type $I$ universe
\begin{equation}\label{25}
\frac{1}{9}\left(\frac{\dot{A}}{A}+\frac{\dot{B}}{B}+\frac{\dot{C}}{C}\right)^{2}
=H_{0}^{2}+\frac{\kappa^{2}\rho_{0}}{3ABC},
\end{equation}
where $H_{0}$ is the Hubble constant having primary implication in
cosmology and $\rho_0$ is an integration constant. The value of
$H_{0}$ corresponds to the rate at which the universe is expanding
today. This equation implies that
\begin{equation}\label{26}
(ABC)^{-1}=\frac{3}{\kappa^{2}\rho_{0}}(H^2-H_{0}^{2}).
\end{equation}
Using EoS in Eq.(\ref{13}), it follows that
\begin{equation}\label{13*}
\frac{\dot{\rho}_{m}}{\rho_{m}}+3H(1+\omega)=0.
\end{equation}
The components of the universe are described by the terms dark
matter and DE. We consider different cases of fluids and their
combination to construct corresponding $F(T)$ models. For example,
for relativistic matter, $\omega=1/3$, for non-relativistic matter,
it is zero and for DE era, it is equal to $-1$ \cite{20}.

\subsubsection*{Case 1 ($\omega=0$):}

This is the case of non-relativistic matter, like cold dark matter
(CDM) and baryons. It is well approximated as pressureless dust and
called the matter dominated era. Inserting $\omega=0$ in
Eq.(\ref{13*}) and using (\ref{26}), we have
\begin{eqnarray}\label{33}
\rho_{m}=\rho_{c}(ABC)^{-1}=\frac{3\rho_{c}}{\kappa^{2}\rho_{0}}(H^2-H_{0}^{2}),
\end{eqnarray}
where $\rho_{c}$ is an integration constant. In terms of torsion scalar, above equation
becomes
\begin{equation}\label{101}
\rho_{m}=\frac{\rho_{c}}{3\kappa^{2}\rho_{0}}(J-9H_{0}^{2}-T).
\end{equation}
Substitution of this value of $\rho_{m}$ in (\ref{11}) implies
that
\begin{equation}\label{102}
2TF_T+F=\frac{2\rho_{c}}{3\rho_{0}}(J-9H_{0}^{2}-T)
\end{equation}
which has the solution
\begin{equation}\label{103}
F(T)=\frac{\rho_{c}}{3\rho_{0}}\left(\frac{1}{\sqrt{T}}\int\frac{J}{\sqrt{T}}dT-18H_{0}^{2}
-\frac{2}{3}T\right).
\end{equation}
This will have a unique solution if the value of $J$ is known which
corresponds to the unknown scale factors. Thus for matter dominated
era, we obtain a model in the form of torsion scalar and Hubble
constant.

\subsubsection*{Case 2 ($\omega=1/3$):}

Here we consider the relativistic matter, like photons and massless
neutrinos, with EoS parameter $\omega=1/3$. This case represents the
radiation dominated era of the universe. Substituting $\omega=1/3$
in Eq.(\ref{13*}) and using (\ref{T}) and (\ref{26}), we obtain
\begin{equation}\label{104}
\rho_{m}=\frac{\rho_{r}}{81\kappa^8\rho_{0}^{4}}(J-9H_{0}^{2}-T)^4,
\end{equation}
where $\rho_r$ is another integration constant. Inserting this
$\rho_m$ in (\ref{11}), we get
\begin{equation}\label{105}
2TF_T+F=\frac{2\rho_{r}}{81\kappa^6\rho_{0}^{4}}(J-9H_{0}^{2}-T)^4
\end{equation}
implying that
\begin{equation}\label{106}
F(T)=\frac{\rho_{r}}{81\kappa^6\rho_{0}^{4}\sqrt{T}}\int\left(\frac{J}{T^{1/8}}-
\frac{9H_{0}^{2}}{T^{1/8}}-T^{1/8}\right)^{4}dT.
\end{equation}
This also depends upon the value of $J$ as well as torsion scalar
and Hubble constant.

\subsubsection*{Case 3 ($\omega=-1$):}

This case represents the present DE constituting $74\%$ of the
universal density. Dark energy is assumed to have a large negative
pressure in order to explain the observed acceleration of the
universe. It is also termed as energy density of vacuum or
cosmological constant $\Lambda$. Replacing $\omega=-1$ in
Eq.(\ref{13*}), we have
\begin{equation}\label{38}
\rho_{m}=\rho_{d},
\end{equation}
where $\rho_{d}$ is an integration constant. Consequently,
Eq.(\ref{11}) takes the form
\begin{equation}\label{107}
2TF_T+F=2\kappa^2\rho_{d}
\end{equation}
with
\begin{equation}\label{39*}
F(T)=2\kappa^2\rho_{d}.
\end{equation}
This turns out to be a constant model which is consistent with
cosmological constant.

\subsubsection*{Case 4 (Combination of $\omega=0$ and $\omega=1/3$):}

Let us now consider the case when the energy density is a
combination of two different fluids, the dust fluid and radiations.
Adding Eqs.(\ref{101}) and (\ref{104}), after simplification, it
follows that
\begin{equation}\label{108}
\rho_{m}=\frac{1}{3\kappa^2\rho_{0}}(J-9H_{0}^2-T)[\rho_c+\frac{\rho_r}
{27\kappa^{6}\rho_{0}^3}(J-9H_{0}^2-T)^3].
\end{equation}
Substituting this $\rho_m$ in (\ref{11}), we get
\begin{equation}\label{109}
2TF_T+F=\frac{2}{3\rho_{0}}(J-9H_{0}^2-T)[\rho_c+\frac{\rho_r}
{27\kappa^{6}\rho_{0}^3}(J-9H_{0}^2-T)^3]
\end{equation}
and hence
\begin{eqnarray}\nonumber
F(T)&=&\frac{\rho_{c}}{3\rho_{0}}\left(\frac{1}{\sqrt{T}}\int\frac{J}{\sqrt{T}}dT-18H_{0}^{2}
-\frac{2}{3}T\right)\\\label{110}
&+&\frac{\rho_{r}}{81\kappa^6\rho_{0}^{4}\sqrt{T}}
\int\left(\frac{J}{T^{1/8}}-\frac{9H_{0}^{2}}{T^{1/8}}-T^{1/8}\right)^{4}dT.
\end{eqnarray}

\subsubsection*{Case 5 (Combination of $\omega=0$ and $\omega=-1$):}

The combination of EoS parameters for matter dominated era and DE
yields
\begin{equation}\label{111}
\rho_{m}=\frac{\rho_c}{3\kappa^2\rho_0}(J-9H_{0}^2-T)+\rho_d.
\end{equation}
Inserting in Eq.(\ref{11}), it follows that
\begin{equation}\label{112}
2TF_T+F=\frac{2\rho_c}{3\rho_0}(J-9H_{0}^2-T)+2\kappa^2\rho_d
\end{equation}
yielding
\begin{equation}\label{113}
F(T)=\frac{\rho_{c}}{3\rho_{0}}\left(\frac{1}{\sqrt{T}}\int\frac{J}{\sqrt{T}}dT-18H_{0}^{2}
-\frac{2}{3}T\right)+2\kappa^2\rho_d.
\end{equation}

\subsubsection*{Case 6 (Combination of $\omega=-1$ and $\omega=1/3$):}

This case gives the following form of the energy density
\begin{equation}\label{114}
\rho_{m}=\rho_d+\frac{\rho_r}{81\kappa^8\rho_0^{4}}(J-9H_{0}^2-T)^4.
\end{equation}
Substituting this value in Eq.(\ref{11}), we get
\begin{equation}\label{115}
2TF_T+F=2\kappa^2\rho_d+\frac{2\rho_r}{81\kappa^6\rho_0^{4}}(J-9H_{0}^2-T)^4
\end{equation}
which gives
\begin{eqnarray}\label{115a}
F(T)=2\kappa^2\rho_d+\frac{\rho_{r}}{81\kappa^6\rho_{0}^{4}\sqrt{T}}\int\left(\frac{J}
{T^{1/8}}-\frac{9H_{0}^{2}}{T^{1/8}}-T^{1/8}\right)^{4}dT.
\end{eqnarray}
It is mentioned here that the cases \textbf{4-6} provide $F(T)$
models for combination of different matters. Normally, the dark
matter and DE developed independently. However, there are attempts
\cite{21} to include an interaction amongst them so that one can get
some insights and see the combined effect of different fluids. Dark
matter plays a central role in galaxy evolution and has measurable
effects on the anisotropies observed in the cosmic microwave
background. Although, matter made a larger fraction of total energy
of the universe but its contribution would fall in the far future as
DE becomes more dominant. It may provide an interaction between dark
matter and DE and can drive transition from an early matter
dominated era to a phase of accelerated expansion. Using the same
phenomenon, DE and different forms of matter are discussed in the
framework of $F(T)$ theory which may help to discuss accelerated
expansion of the universe.

\subsection{Using EoS Parameter}

Here we formulate some $F(T)$ models in a slightly different way. We
substitute different values of parameter $\omega$ in Eq.(\ref{20})
and solve it accordingly. Equation (\ref{20}) can be written as
\begin{equation}\label{47}
4ZF_{TT}+[-4U(\omega+1)+2E-4Y]F_{T}+(\omega+1)F=0.
\end{equation}
Now we construct $F(T)$ models in the following cases.

\subsubsection*{Case 1:}

Putting $\omega=\frac{1}{3}$ in Eq.(\ref{47}), we
have
\begin{equation}\label{48}
2ZF_{TT}+(E-2Y-\frac{8}{3}U)F_{T}+\frac{2}{3}F=0.
\end{equation}
This has the following general solution
\begin{eqnarray}\nonumber
F(T)&=&c_{3}\exp
T\left[\frac{8U-3E+6Y+\sqrt{(3E-8U-6Y)^{2}-48Z}}{12Z}\right]\\\label{49}
&+&c_{4}\exp
T\left[\frac{8U-3E+6Y-\sqrt{(3E-8U-6Y)^{2}-48Z}}{12Z}\right],
\end{eqnarray}
where $c_3$ and $c_4$ are constants.

\subsubsection*{Case 2:}

Consider the case when pressure is zero, i.e, $\omega=0$. In this
case, Eq.(\ref{47}) implies
\begin{equation}\label{50}
4ZF_{TT}+[-4U+2E-4Y]F_{T}+F=0.
\end{equation}
It has the following general solution
\begin{eqnarray}\nonumber
F(T)&=&c_{5}\exp
T\left[\frac{6U-3E+6Y+\sqrt{(-6U+3E-6Y)^{2}-24Z}}{12Z}\right]\\\label{51}
&+&c_{6}\exp
T\left[\frac{6U-3E+6Y-\sqrt{(-6U+3E-6Y)^{2}-24Z}}{12Z}\right],
\end{eqnarray}
where  $c_5$ and $c_6$ are arbitrary constants.

\subsubsection*{Case 3:}

For $\omega=-1$, Eq.(\ref{47}) becomes
\begin{equation}\label{52}
4ZF_{TT}+(2E-4Y)F_{T}=0
\end{equation}
which has two solutions. The first solution yields $F(T)=c_{7}$
while the second solution is given by
\begin{eqnarray}\label{54}
F(T)=c_{8}\exp\left[\left(\frac{2Y-E}{2Z}\right)T\right],
\end{eqnarray}
where $c_7$ and $c_8$ are constants. Equations (\ref{49}),
(\ref{51}) and (\ref{54}) represent $F(T)$ models corresponding to
radiation, matter and DE phases respectively. The exponential form
of $F(T)$ models represents a universe which always lies in phantom
or non-phantom phase depending on parameters of the models
\cite{22}.

\section{Construction of EoS Parameter and Cosmic Acceleration}

In this section, we derive EoS parameter by using two different
$F(T)$ models and also investigate cosmic acceleration. For this
purpose, we evaluate $\rho_{m}$ and $p_{m}$ using the field
equations and then construct the corresponding EoS parameter.

\subsection{The First Model}

Consider the following $F(T)$ model \cite{4}
\begin{equation}\label{62}
F=\alpha T+\frac{\beta}{T},
\end{equation}
where $\alpha$ and $\beta$ are positive real constants. Replacing
this value of $F$ in Eqs.(\ref{11}) and (\ref{12}), it follows that
\begin{eqnarray}\label{63}
2\kappa^{2}\rho_{m}&=&(-4U+T)\alpha+\beta(1+4UT^{-1})T^{-1},\\\label{64}
2\kappa^{2}p_{m}&=&(2E-T)\alpha-\beta(1+2ET^{-1}+8ZT^{-2})T^{-1}.
\end{eqnarray}
Dividing Eq.(\ref{64}) by (\ref{63}), the EoS parameter is obtained
as follows
\begin{equation}\label{65}
\omega=-1+\frac{2(Y-U)(\alpha-\beta T^{-2})-8Z\beta
T^{-3}}{(-4U+T)\alpha+ \beta (1+4UT^{-1})T^{-1}}.
\end{equation}
Now we discuss this equation for particular values of $\alpha$ and
$\beta$. For $\alpha\neq0,~\beta=0$, we obtain
\begin{equation}\label{65*}
\omega=-1+\frac{1}{3}\left(1-\frac{Y}{U}\right).
\end{equation}
This leads to three different cases of $\omega$ representing
different phases of the evolution of the universe as follows:
\begin{itemize}
\item If $\frac{Y}{U}>1$, then $\omega<-1$ which corresponds to
the phantom accelerating universe.
\item When $\frac{Y}{U}<1$, the EoS
parameter will be slightly greater than $-1$ which means that the
universe stays in the quintessence region.
\item If $\frac{Y}{U}=1$, we obtain a universe whose dynamics
is dominated by cosmological constant with $\omega=-1$.
\end{itemize}
It is interesting to mention here that model (\ref{65*}) reduces to
GR spatially flat Friedmann equation in the limiting case when
anisotropy vanishes. The case $\alpha=0,~\beta\neq0$ does not
provide meaningful results.

\subsection{The Second Model}

Assume $F(T)$ has the form \cite{4}
\begin{equation}\label{66}
F=\alpha T+\beta T^{n},
\end{equation}
where $n$ is a positive real number. The corresponding field
equations will become
\begin{eqnarray}\label{67}
2\kappa^{2}\rho_{m}&=&(-4U+T)\alpha+\beta(-4UnT^{-1}+1)T^{n},\\
\label{68}2\kappa^{2}p_{m}&=&(2E-T)\alpha +2En\beta
T^{n-1}-4Z\beta n(n-1) T^{n-2}-\beta T^{n}.
\end{eqnarray}
Consequently, the EoS parameter takes the form
\begin{equation}\label{69}
\omega=-1-\frac{2(U-Y)\alpha+2n\beta(U-Y) T^{n-1}+4Z\beta n(n-1)
T^{n-2}}{(-4U+T)\alpha+\beta (-4Un T^{-1}+1)T^{n}}.
\end{equation}
The case $\alpha\neq0,~\beta=0$ leads to the same discussion as in
the first case. For $\alpha=0,~\beta\neq0$, we have
\begin{equation}\label{69*}
\omega=-1+\frac{n}{2n+1}\left[1-\left\{\frac{Y}{U}+\frac{Z(n-1)}{U^2}\right\}\right].
\end{equation}
For any positive real number $n$, we can discuss as follows:
\begin{itemize}
\item When ${\frac{Y}{U}+\frac{Z(n-1)}{U^2}}>1$, Eq.(\ref{69*}) gives $\omega<-1$
 which represents the
phantom accelerating universe.
\item For ${\frac{Y}{U}+\frac{Z(n-1)}{U^2}}=1$, we obtain $\omega=-1$ and hence
the universe rests in DE era dominated by cosmological constant.
\item The case ${\frac{Y}{U}+\frac{Z(n-1)}{U^2}}<-1$ corresponds to the
quintessence era because $\omega>-1$.
\end{itemize}

Assuming $n=1$ as a particular case in Eqs.(\ref{67}) and
(\ref{68}), we have
\begin{eqnarray}\label{73}
\rho_{m}&=&\frac{1}{2\kappa^{2}}(\alpha+\beta)(-4U+T),\\
\label{73*}p_{m}&=&\frac{1}{2\kappa^{2}}(\alpha+\beta)(2E-T).
\end{eqnarray}
In the following, we discuss the evolution of the scale factor for
Bianchi type $I$ universe. For this purpose, we assume \cite{4}
\begin{equation}\label{71}
p_{m}=\frac{A_{-1}(T)}{\rho_{m}}+A_{0}(T)+A_{1}(T)\rho_{m}
\end{equation}
such that $A_{-1},~A_0,~A_1$ are constants. Substituting Eqs.(\ref{73})
and (\ref{73*}) in the above equation, it follows that
\begin{equation}\label{74}
2E-T=\frac{a}{-4U+T}+b+c(-4U+T),
\end{equation}
where
\begin{equation}\label{75}
a=\frac{4\kappa^{4}A_{-1}}{(\alpha+\beta)^{2}},\quad
b=\frac{2\kappa^{2}A_{0}}{\alpha+\beta},\quad c=A_{1}.
\end{equation}
This equation leads to
\begin{eqnarray}\label{76}
T&=&\frac{4U+2E-b+8Uc}{2(1+c)}\pm\frac{1}{2(1+c)}\nonumber\\
&\times&[b^{2}-4a-4ac-4Eb+4E^{2}+8Ub-16EU+16U^{2}]^{1/2}.
\end{eqnarray}
Substituting this value of torsion in Eq.(\ref{T'}), we have
\begin{eqnarray}\label{78}
H&=&\frac{1}{3}\left[\left|J-\frac{4U+2E-b+8Uc}{2(1+c)}\mp\frac{1}{2(1+c)}\right.\right.
\nonumber\\
&\times&\left.\left.\sqrt{b^{2}-4a-4ac-4Eb+4E^{2}+8Ub-16EU+16U^{2}}\right|\right]^{1/2}.
\end{eqnarray}
The correspondingly average scale factor becomes
\begin{eqnarray}\label{80}
R&=&R_{0}\exp\left\{\frac{1}{3}\int\left[\left|J-\frac{4U+2E-b+8Uc}{2(1+c)}
\mp\frac{1}{2(1+c)}\right.\right.
\right.\nonumber\\&\times&\left.\left.\left.\sqrt{b^{2}-4a-4ac-4Eb+4E^{2}+8Ub-16EU+16U^{2}}
\right|\right]^{1/2}dt\right\}.
\end{eqnarray}

As a special case of model (\ref{71}), if we take $A_{-1}$ as a
constant while $A_0=0=A_1$, we obtain standard Chaplygin gas EoS
\cite{23}. In this respect, Eqs.(\ref{76}) and (\ref{78}) give the
following results respectively
\begin{eqnarray}\label{81}
T&=&(E+2U)\pm\sqrt{(E-2U)^2-a}~,\\\label{82}
H&=&\frac{1}{3}\left[J-(E+2U)\mp\sqrt{(E-2U)^2-a}\right] .
\end{eqnarray}
The average scale factor for chaplygin gas has the form
\begin{equation}\label{83}
R=R_0\exp\left\{\frac{1}{3}\int\left[\left|J-(E+2U)\mp\sqrt{(E-2U)^2-a}\right|\right]
dt\right\}.
\end{equation}
This represents an exponential expansion which may result a rapid
increment between the distance of two non-accelerating observers
as compared to the speed of light. As a result, both observers are
unable to contact each other. Thus if our universe is forthcoming
to a de Sitter universe \cite{5}, then we would not be able to
observe any galaxy other than our own Milky way system.

\section{Summary}

In this paper, we have investigated the recently developed $F(T)$
gravity, where $T$ is responsible for the cosmic acceleration
without DE component. For this purpose, we have taken Bianchi type
$I$ spacetime which is one of the simplest models describing
anisotropic, spatially homogenous and flat universe. Some $F(T)$
gravity models have been constructed by using two different
approaches. In the first approach, we have used the continuity
equation while in the second method, EoS parameter is used. These
$F(T)$ gravity models represent three different phases of the
universe exhibiting different values of EoS parameter. The matter,
radiation and DE eras respectively correspond to
$\omega=0,~\omega=1/3$ and $\omega=-1$.

Matter dominated era describes expansion of the universe filled with
non-interacting dust particles while radiation dominated era
represents early universe after the hot big bang. The DE case
corresponds to the universe dominated by a strong negative pressure
causing late-time acceleration. If we consider combination of
radiation and matter, we may have more interesting results to study
the developing universe. We have also constructed some models by
using different combinations of EoS parameter. Also, we have
obtained $F(T)$ models in exponential form for some particular
values of EoS parameter.

Since the evolution of EoS parameter is one of the biggest efforts
in observational cosmology today. We have considered two
well-known $F(T)$ models and found the corresponding expression
for $\omega$. These have been discussed for some particular values
of parameters $\alpha,~\beta$ which yield fruitful results
corresponding to realistic situations. The cosmic acceleration has
been discussed by using some $F(T)$ models. We conclude that our
universe would approach to de Sitter universe in the infinite
future. It is interesting to mention here that one of the $F(T)$
models (case 3) inherits a constant solution which may correspond
to the cosmological constant. Notice that the isotropic expansion
of the universe is obtained for $\Delta = 0$ which depends upon
the values of unknown scale factors and parameters involved in the
corresponding models \cite{S20}-\cite{S22}. It is worthwhile to
point out here that our results correspond to FRW universe for the
special case, i.e., $A(t)=B(t)=C(t)=a(t)$.

\end{document}